\documentstyle[prl,aps,psfig]{revtex}
\newlength{\defaultparindent}
\begin{document}

\draft

\title{An Improved Determination of $\alpha _{S}$ From Neutrino-Nucleon Scattering}

\author{W.~G.~Seligman,$^{2}$ C.~G.~Arroyo,$^{2}$ L.~de~Barbaro,$^{5}$ P.~de~Barbaro,%
$^{6}$ A.~O.~Bazarko,$^{2}$ R.~H.~Bernstein,$^{3}$ A.~Bodek,$^{6}$ T.~Bolton,%
$^{4}$ H.~Budd,$^{6}$ J.~Conrad,$^{2}$ D.~A.~Harris,$%
^{6} $ R.~A.~Johnson,$^{1}$ J.~H.~Kim,$^{2}$ B.~J.~King,$^{2}$ T.~Kinnel,$%
^{7}$ M.~J.~Lamm,$^{3}$ W.~C.~Lefmann,$^{2}$ W.~Marsh,$^{3}$ K.~S.~McFarland,%
$^{3}$ C.~McNulty,$^{2}$ S.~R.~Mishra,$^{2}$ D.~Naples,$^{4}$ P.~Z.~Quintas,$%
^{2}$ A.~Romosan,$^{2}$ W.~K.~Sakumoto,$^{6}$ H. Schellman,$^{5}$
F.~J.~Sciulli,$^{2}$ M.~H.~Shaevitz,$^{2}$ W.~H.~Smith,$^{7}$ P.~Spentzouris,%
$^{2}$ E.~G.~Stern,$^{2}$ M.~Vakili,$^{1}$ U.~K.~Yang,$^{6}$ and J.~Yu$^{3}$}

\address{$^{1}$ University of Cincinnati, Cincinnati, OH 45221 \\
$^{2}$ Columbia University, New York, NY 10027 \\
$^{3}$ Fermi National Accelerator Laboratory, Batavia, IL 60510 \\
$^{4}$ Kansas State University, Manhattan, KS 66506 \\
$^{5}$ Northwestern University, Evanston, IL 60208 \\
$^{6}$ University of Rochester, Rochester, NY 14627 \\
$^{7}$ University of Wisconsin, Madison, WI 53706 }

\date{12 May 1997}

\maketitle

\begin{abstract}
We present an improved determination of the proton structure functions
$F_{2}$ and $xF_{3}$ from the CCFR $\nu $-Fe deep inelastic scattering
(DIS) experiment.  Comparisons to high-statistics charged-lepton
scattering results for $F_{2}$ from the NMC, E665, SLAC, and BCDMS
experiments, after correcting for quark-charge and heavy-target
effects, indicate good agreement for $x>0.1$ but some discrepancy at
lower $x$. The $Q^{2}$ evolution of both the $F_{2}$ and $xF_{3}$
structure functions yields the quantum chromodynamics (QCD) scale
parameter $\Lambda _{\overline{MS}}^{NLO,(4)}=337 \pm 28$(exp.)$\
MeV$. This corresponds to a value of the strong coupling constant at
the scale of mass of the $Z$-boson of $\alpha _{S}(M_{Z}^{2})=0.119
\pm 0.002$(exp.)$\pm 0.004$(theory) and is one of the most precise
measurements of this quantity.

\end{abstract}



High-energy neutrinos are a unique probe for testing QCD and understanding
the parton properties of nucleon structure. Combinations of neutrino and
antineutrino scattering data are used to determine the $F_{2}$ and $xF_{3}$
structure functions (SFs) which determine the valence, sea, and gluon parton
distributions in the nucleon \cite{GRVMRS,CTEQ}. The universalities of
parton distributions can also be studied by comparing neutrino and
charged-lepton scattering data. Past measurements have indicated that 
$F_{2}^{\nu }$ differs from $F_{2}^{e/\mu }$ by 10-20\% in the low-$x$ region.
These differences are larger than the quoted experimental errors of the
measurements and may indicate the need for modifications of the theoretical
modeling to include higher-order or new physics contributions. QCD predicts
the scaling violations ($Q^{2}$ dependence) of $F_{2}$ and $xF_{3}$ and,
experimentally, the observed scaling violations can be tested against those
predictions to determine $\alpha _{S}$ \cite{DGLAP} or the related QCD scale
parameter, $\Lambda _{QCD}$. The $\alpha _{S}$ determination from neutrino
scattering has a small theoretical uncertainty since the electroweak
radiative corrections, scale uncertainties, and next-to-leading order (NLO)
corrections are well understood. 

In this paper, we present an updated
analysis of the Columbia-Chicago-Fermilab-Rochester (CCFR) collaboration
neutrino scattering data with improved estimates of quark model parameters 
\cite{SAR} and systematic uncertainties. The $\alpha _{S}$ measurement from
this analysis is one of the most precise due to the high energy and statistics of the
experiment compared to previous measurements \cite{CDHSW,Oltman}.

The differential cross sections for the $\nu $-$N$ charged-current process 
$\nu _{\mu }\left( \overline{\nu}_{\mu }\right) +N\rightarrow \mu ^{-}\left(
\mu ^{+}\right) +X$, in terms of the Lorentz-invariant structure functions 
$F_{2} $, $2xF_{1}$, and $xF_{3}$ are: 
\begin{eqnarray}
\frac{d\sigma ^{\nu ,\overline{\nu }}}{dx\ dy} &=& \frac{G_{F}^{2}ME_{\nu }}{
\pi }\left[ \left( 1-y-\frac{Mxy}{2E_{\nu }}\right) F_{2}\left(
x,Q^{2}\right) \right. \nonumber \\
&&\left. +\frac{y^{2}}{2}2xF_{1}\left( x,Q^{2}\right) \pm y\left( 1-%
\frac{y}{2}\right) xF_{3}\left( x,Q^{2}\right) \right]  \label{eq:dxdy}
\end{eqnarray}
where $G_{F}$ is the weak Fermi coupling constant, $M$ is the nucleon
mass, $E_{\nu }$ is the incident neutrino energy, $Q^{2}$ is the
square of the four-momentum transfer to the nucleon, the scaling
variable $y=E_{HAD}/E_{\nu }$ is the fractional energy transferred to
the hadronic vertex with $E_{HAD}$ equal to the measured hadronic
energy, and $x=Q^{2}/2ME_{\nu }y$, the Bjorken scaling variable, is
the fractional momentum carried by the struck quark. The structure
function $2xF_{1}$ is expressed in terms of $F_{2}$ by
$2xF_{1}(x,Q^{2})=F_{2}(x,Q^{2})\times
\frac{1+4M^{2}x^{2}/Q^{2}}{1+R(x,Q^{2})}$, where $R=\frac{\sigma
_{L}}{\sigma _{T}} $ is the ratio of the cross-section of
longitudinally to transversely-polarized $W$-bosons. In the
leading-order quark-parton model, $F_{2}$ is the singlet distribution
$xq^{S}=x\sum \left( q+\overline{q}\right) $, the sum of the momentum
densities of all interacting quark constituents, and $xF_{3}$ is the
non-singlet distribution $xq^{NS}=x\sum \left( q-\overline{q}\right)
=xu_{V}+xd_{V}$, the valence quark momentum density; these relations
are modified by higher-order QCD corrections.

The neutrino DIS data were taken in two high-energy high-statistics
runs, FNAL E744 and E770, in the Fermilab Tevatron fixed-target
quadrupole triplet beam (QTB) line by the CCFR collaboration. The
detector, described in Refs.~\cite{hadcal,mucal}, consists of a target
calorimeter instrumented with both scintillators and drift chambers
for measuring the energy of the hadron shower $E_{HAD}$ and the muon
angle $\theta _{\mu }$, followed by a toroid spectrometer for
measuring the muon momentum $p_{\mu }$. There are 950,000 $\nu _{\mu
}$ events and 170,000 $\overline{\nu }_{\mu }$ events in the data
sample after fiducial-volume cuts, geometric cuts, and kinematic cuts
of $p_{\mu }$ $>15\ GeV$, $\theta _{\mu }<150\ mr$, $E_{HAD}$ $>10\
GeV$, and $30<E_{\nu }<360\ GeV$, to select regions of high efficiency
and small systematic errors in reconstruction.

In order to calculate the SF in Eq.~(\ref{eq:dxdy}) from the number of
observed $\nu _{\mu }$ and $\overline{\nu }_{\mu }$ events, it is
necessary to determine the $\nu _{\mu }$ and $\overline{\nu }_{\mu }$
flux.  No direct measurement of the absolute flux was possible in the
QTB. The absolute normalization of the $\nu _{\mu }$ flux was fixed to
the constraint that the neutrino-nucleon total cross-section equaled
the world average of the isoscalar-corrected Fe target experiments,
$\sigma ^{\nu \,Fe}/E_{\nu }=(0.677\pm 0.014)\times
10^{-38}cm^{2}/GeV$ \cite{Auc,worldsig}. The relative flux
determination, i.e.  the ratio of the flux between different energies
and between $\nu _{\mu }$ and $\overline{\nu }_{\mu }$, was determined
from the low- $E_{HAD}$ events using a technique described in
Ref.~\cite{WGSthesis,Auc,BelRein}.  The cross-sections, multiplied by
the flux, are compared to the observed number of $\nu $-$N$ and
$\overline{\nu }$ -$N$ events in an $x$ and $Q^{2}$ bin to extract
$F_{2}(x,Q^{2})$ and $xF_{3}(x,Q^{2})$.

\begin{figure}
\centerline{\psfig{figure=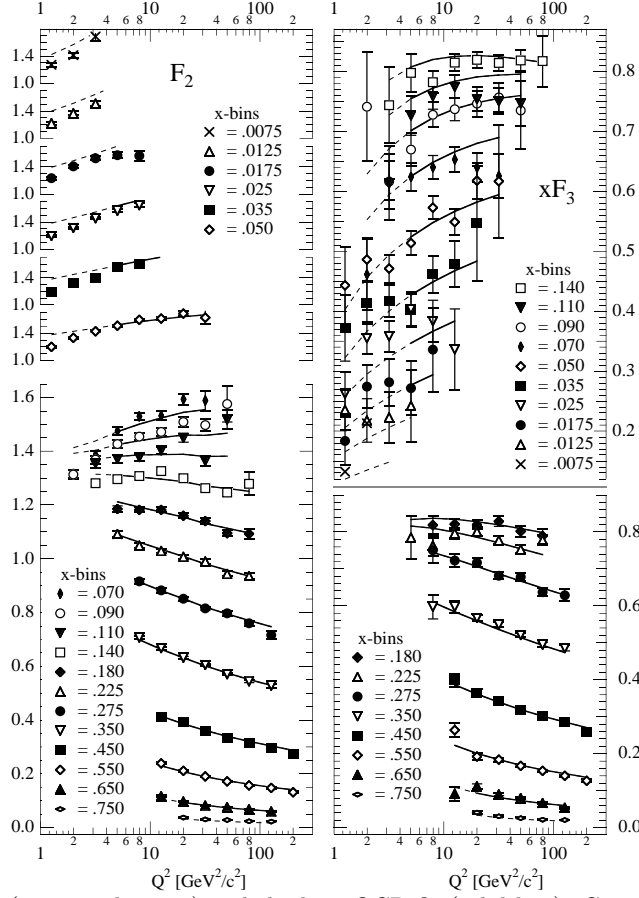,width=3.375in}}
\caption{The $F_{2}$ and $xF_{3}$ data (statistical errors) and the best QCD
fit (solid line). Cuts of $Q^{2}>5\ GeV^{2}$, $W^{2}>10\ GeV^{2}$, and $x<0.7
$ were applied for the NLO-QCD fit which include target mass corrections. The
dashed line extrapolates the QCD fit into the data regions excluded by the
cuts. Deviations of the data from the extrapolated fit are partly
due to non-perturbative effects.}
\label{fig:SF}
\end{figure}

SFs extracted from the CCFR data have been previously presented
\cite{PZQprl}.  In the earlier analysis, the muon and hadron energy
calibrations were determined using a Monte Carlo technique in an
attempt to reduce the dominant source of systematic error in the
analysis, the relative calibration between the muon and hadron
energies. 
Our subsequent analysis determined that the control of systematic
errors for this technique was insufficient to justify its continued use.
This paper presents a re-extraction of the
SFs that uses the calibrations directly determined from the test beam
data collected during the course of the experiment
\cite{hadcal,mucal}, which results in a net change of +2.1\% in the
relative calibration and an increase in the corresponding systematic
error to 1.4\%.  Other changes in the SF extraction include more
complete radiative corrections \cite{radcorr}, and the value of $R$
now used in the extraction comes from a global fit to the world's
measurements \cite {Rworld}. In addition, the estimates of the
experimental and theoretical systematic errors in the analysis are
improved \cite {WGSthesis}. The structure functions are corrected for
radiative effects \cite{radcorr}, the non-isoscalarity of the Fe
target, the charm-production threshold \cite{Bar76,TMC}, and the mass
of the $W$-boson propagator.  The SFs with statistical errors, along
with the QCD fits described below, are shown in Fig.~\ref{fig:SF}
\cite{URL}.

The structure function $F_{2}$ from $\nu $ DIS on iron can be compared
to $F_{2}$ from $e$ and $\mu $ DIS on isoscalar targets. To make this
comparison, two corrections must be made to the charged-lepton
data. For deuterium data, a heavy nuclear target correction must be
made to convert $F_{2}^{\ell D}$ to $F_{2}^{\ell \,Fe}$. This
correction was made by parameterizing the $F_{2}^{\ell N}/F_{2}^{\ell
D}$ data from SLAC and NMC \cite{heavytarget}. $F_{2}$ from electromagnetic
interactions couples to the constituent quarks with the square of the
quark electric charge. Thus a second correction is necessary:
\begin{equation}
\frac{F_{2}^{\ell }}{F_{2}^{\nu }} = \frac{5}{18}\left( 1-\frac{3}{5}\frac{%
\left( s+\overline{s}-c-\overline{c}\right) }{\left( q+\overline{q}\right) }
\right)  \label{eq:five18ths}
\end{equation}

\begin{figure}[tbp]
\centerline{\psfig{figure=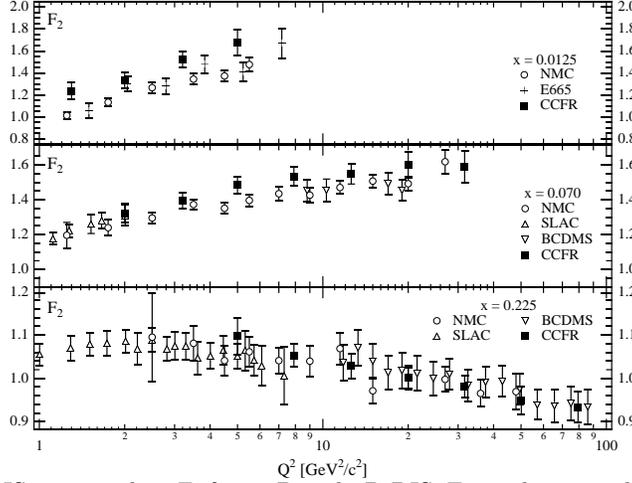,width=3.375in}}
\caption{$F_{2}$ from CCFR $\nu $-Fe DIS compared to $F_{2}$ from $e$D and 
$\mu $D DIS. Errors bars are the statistical and systematic errors added in
quadrature. The charged-lepton data have been corrected to an isoscalar Fe
target, and for quark-charge effects in the DIS
scheme which is valid to all orders (see text). The
NMC data plotted in the figure were extracted with the same $R$ as used in
the CCFR analysis \protect\cite{NMC}.}
\label{fig:compare}
\end{figure}

This formula is exact to all orders in QCD in the DIS renormalization
scheme, so for the purposes of this comparison the charged-lepton
structure functions were corrected according to
Eq.~(\ref{eq:five18ths}) with quark distributions given by CTEQ4D
\cite{CTEQ}, which parameterizes the parton densities in the DIS
scheme. The errors on the nuclear and charge corrections are small
compared to the statistical and systematic errors on both the CCFR and
NMC data. The corrected structure functions from NMC, E665, SLAC, and
BCDMS \cite{NMC,F2l} for selected $x$-bins are shown in
Fig.~\ref{fig:compare}. There is a 15\% discrepancy between the NMC
charged-lepton $F_{2}$ and the CCFR neutrino $F_{2}$ at $x=0.0125$. As
the value of $x$ increases, the discrepancy decreases, until there is
agreement between CCFR and the charged-lepton experiments above
$x=0.1$.

The discrepancy between CCFR and NMC at low $x$ is outside the
experimental systematic errors quoted by the groups and several
suggestions for an explanation have been put forward. One suggestion
\cite{CTEQ1}, that the discrepancy can be entirely explained by a
large strange sea, is excluded by the CCFR dimuon analysis which
directly measures the strange sea \cite{AOB}. Other suggestions are
that the strange sea may have a different distribution than the
normally-assumed form \cite{ssbar}, or that the heavy nuclear target
correction may be different between neutrinos and charged
leptons. More experimental data will be necessary to resolve this
issue.

According to perturbative QCD (PQCD), the $Q^{2}$ dependence of the
quark momentum densities is described by ``evolution equations''
\cite{DGLAP}.  The evolution of the non-singlet distribution does not
depend on assumptions about the gluons, but the singlet distribution
co-evolves with the gluon distribution. The previous CCFR analysis
\cite{PZQprl} only compared the SF to the non-singlet evolution. This
analysis takes advantage of the ability of neutrino DIS to measure
both $F_{2}$ and $xF_{3}$, and simultaneously evolves the non-singlet,
singlet, and gluon distributions for a more precise determination of
$\Lambda _{QCD}$.

Systematic uncertainties in the structure function extraction were
investigated, leading to correlated errors for each of the data points
in Fig.~\ref{fig:SF}. The largest sources of systematic error in the
determination of $\Lambda _{QCD}$ are the muon and hadron absolute
energy calibrations. The error in the energy calibration was measured
to be 1\% for $p_{\mu }$ \cite{mucal}, and 1\% for $E_{HAD}$ for the
E744 and E770 data separately \cite{hadcal}. Another major source of
systematic error is the error in the value of $\sigma ^{\overline{\nu
} }/\sigma ^{\nu }$, the ratio of the total $\overline{\nu }$ to $\nu
$ cross-section. The value chosen was the world average of $\nu $-Fe
DIS experiments, including this one \cite{worldsig,WGSthesis}, $\sigma
^{\overline{\nu }}/\sigma ^{\nu }=0.499\pm 0.007$. Other sources of
systematic error were investigated, including systematic errors in the
flux extraction and variations in the physics model used in the Monte
Carlo, but the effects of these other sources were small
\cite{WGSthesis}.  To determine the uncertainty for each source, the
structure functions $F_{2}$ and $xF_{3}$ are extracted with the given
systematic quantity changed by one error unit up and down, where an
``error unit'' is the best estimate of the systematic error prior to
the fit described below.  The difference of these modified structure
functions and the standard ones gives the point-to-point correlated
systematic errors in $F_{2}$ and $xF_{3}$ for each $(x,Q^{2})$
bin. Complete tables of errors can be found in
Ref.~\cite{WGSthesis,URL}.

For the PQCD analysis of the structure functions, we performed a $\chi
^{2}$ fit which minimizes the difference between a theoretical
prediction and the measured values of $F_{2}$ and $xF_{3}$ in each
$(x,Q^{2})$ bin. The theoretical prediction is obtained using the Duke
and Owens NLO QCD evolution program \cite{DOprog,WGSthesis}. The
prediction incorporates a parameterization of the parton distributions
for the singlet, non-singlet, and gluon distributions at a reference
value $Q_{0}^{2}=5\,GeV^{2}$ as shown in Table \ref{tab:fitresults}
and includes $\Lambda _{NLO}$ as a fit parameter. The prediction is
compared to the structure function data using a $\chi ^{2}$ that
includes the statistical errors (including the $\Delta F_{2}\Delta
xF_{3}$ correlations) and the correlated systematic uncertainties. The
systematic errors are included by introducing a parameter $\delta
\left( k\right) $ for each systematic uncertainty. This parameter
controls the amount of systematic deviation added to the structure
function and is also included in the $\chi ^{2}$ function
(Eq.~(\ref{eq:chi2})). For this procedure, we define the
structure-function vector $\vec{F}$ = $\left(
\begin{array}{cc}
F_{2} & xF_{3}
\end{array}
\right) ^{T}$ and the structure-function statistical error matrix $\widehat{V%
}=\left( \sigma _{ij}\right) $ for $i,j=\{F_{2},xF_{3}\}$. Then the $\chi
^{2}$ for a global fit is given by: 
\begin{equation}
\vec{F}^{diff}=\vec{F}^{data}-\vec{F}^{theory}+\sum\limits_{k}\delta \left(
k\right) \left( \vec{F}^{k}-\vec{F}^{data}\right)  \label{eq:fdiff}
\end{equation}
\begin{equation}
\chi ^{2}=\left( \vec{F}^{diff}\right) ^{T}\ \widehat{V}\ \left( \vec{F}%
^{diff}\right) +\sum\limits_{k}\delta \left( k\right) ^{2}  \label{eq:chi2}
\end{equation}
where $\vec{F}^{data}$ are the measured values as shown in
Fig.~\ref{fig:SF}, $\vec{F}^{theory}$ are the predictions from the
evolution program that depend on fit parameters including $\Lambda
_{\overline{MS}}$, and $\vec{F}^{k}$ are the structure functions
measured with the $k$-th systematic uncertainty changed by one
standard error.

The effects of target mass \cite{TMC} were included in the
fit. Calculations of the effects of higher-twist terms (HT) have
recently become available \cite{DW} and are in agreement with the
measurements of the $F_{2}$ HT \cite{VM}. However, the data in
Ref.~\cite{VM} were analyzed with a value of $\alpha _{S}$ smaller
than our present value which would increase the measurement of HT. An
analysis of HT from preliminary CCFR $xF_{3}$ data \cite{Sidorov}
indicates that the calculation of Ref.~\cite{DW} yields HT that are
too large. For this analysis, the values of the $F_{2}$ and $xF_{3}$
HT corrections were taken to be one-half the values from
Ref.~\cite{DW}, with a conservative systematic error given by
repeating the fit with no HT correction and with the full HT from
Ref.~\cite{DW}.

Cuts of $Q^{2}>5\ GeV^{2}$ and the invariant mass-squared of the
hadronic system $W^{2}>10\ GeV^{2}$ were applied to the data to
include only the perturbative region, and an $x<0.7$ cut includes the
$x$-bins where the resolution corrections are insensitive to Fermi
motion. The $E_{\nu }<360\ GeV$ cut implies an effective limit of
$Q^{2}< 125\ GeV^{2}$. The best QCD fits to the data are shown in
Fig.~\ref{fig:SF}, and the results of the fit are shown in Table
\ref{tab:fitresults}. The $\delta \left( k\right) $ values from the
fit are all zero within two standard deviations, and have errors that
range from 0.12 to 0.98. The fact that these errors are all less than
1.0 indicates that the data coupled with the theory of QCD forms a
more restrictive constraint on the systematic error than the
variations described above.

From this fit, we obtain a measured value of $\Lambda _{\overline{MS}}$ in
NLO QCD for 4 quark flavors of $337\pm 28$(exp.)$\pm 13$(HT)$\ MeV$, which yields $\alpha
_{S}(M_{Z}^{2})=0.119\pm 0.002$(exp.)$\pm 0.001$(HT)$\pm 0.004$(scale), 
where the
error due to the renormalization and factorization scales comes
from Ref.~\cite{VM}. 
The fit also yields a measurement of the
gluon distribution $xG(x,Q_{0}^{2}=5\ GeV^{2})=\left( 2.22\pm 0.34\right)
\times \left( 1-x\right) ^{4.65\pm 0.68}$ in the region $0.04<x<0.70$ which is
consistent with gluon distributions given in Refs.~\cite{GRVMRS,CTEQ}.
A fit to only the $xF_{3}$ data, which is not
coupled to the gluon distribution, gives $\Lambda _{\overline{MS}}
=381\pm 53$(exp.)$\pm 17$(HT)$\ MeV$, which is consistent with the result
of the combined $F_{2}$ and $xF_{3}$ fit but has larger errors because effectively only
half the data are used.
If the systematic uncertainties are not allowed to vary in the
$F_{2}$ and $xF_{3}$ fit and all effects of systematic 
uncertainties are added in quadrature, the value of 
$\Lambda _{\overline{MS}}$ is found to be 
$381\pm 23$(stat.)$\pm 58$(syst.)$\ MeV$.

This result is higher than our
previous measurement \cite{PZQprl}, $\alpha _{S}(M_{Z}^{2})=
0.111\pm 0.002$(stat.)$\pm 0.003$(syst.),
mainly due to effects of the new energy calibrations. 
The current measurement is also larger than the
muon DIS result by the SLAC/BCDMS collaboration \cite{VM}, $\alpha
_{S}(M_{Z}^{2})=0.113\pm 0.003$(exp.)$\pm 0.004$(theory); note that this
theoretical error and the CCFR theory error are from the same calculation.
The low-$x$ discrepancy in $F_{2}$ between CCFR and NMC has a negligible effect
on the $\alpha _{S}$ measurement which is derived mainly from the high-$x$ data.

In summary, a comparison of $F_{2}$ from $\nu $ DIS to that from
charged-lepton DIS shows good agreement above $x=0.1$
but a difference at smaller $x$ that grows
to 15\% at $x\approx 0.01$. We have presented a new, high-precision
measurement of $\Lambda _{\overline{MS}}=337\pm 28\ MeV$ from a
fit to the simultaneous $Q^{2}$ evolution of $F_{2}$ and $xF_{3}$. This
corresponds to a value of
$\alpha _{S}(M_{Z}^{2})=0.119\pm 0.002$(exp.)$\pm 0.004$(theory) and is the most
precise DIS measurement of this quantity.

\begin{table}[tbp]
\caption{Results of the global systematic fit to the CCFR data. The parton
distributions at $Q_{0}^{2}=5\ GeV^{2}$ are parameterized by $%
xq^{NS}\left( x\right) =A_{NS}x^{\eta _{1}}\left( 1-x\right) ^{\eta _{2}}$, $%
xq^{S}\left( x\right) =xq^{NS}\left( x\right) +A_{S}(1-x)^{\eta _{S}},$ 
$xG\left( x\right) =A_{G}\left( 1-x\right) ^{\eta
_{G}}.$ $\delta \left( k\right) $ is the fractional shift for the best value
of systematic quantity $k$ as determined by the fit. Only the most important sources of
systematic error are shown. $\delta \left(
C_{HAD}^{744}\right) $ is the shift for the E744 hadron energy calibration,
$\delta \left( C_{HAD}^{770}\right) $ is the shift for the E770 hadron energy
calibration, $\delta \left( C_{\mu }\right) $ is the shift for the muon
energy calibration, and $\delta \left( \sigma ^{\overline{\nu }}/\sigma
^{\nu }\right) $ is the shift for the ratio of the total $\overline{\nu }$
to $\nu $ cross-section. The $\chi ^{2}$ of the fit is 158 for 164 degrees of freedom.}
\begin{center}
\begin{tabular}{l|r||l|r}                                 
Parameter                             & Fit Results        
  & Parameter                             & Fit Results        \\ \hline 
$\Lambda _{\overline{MS}}$            & $337\pm 28\ MeV$   
  & $A_{G}$                               & $2.22\pm 0.34$     \\ 
$\eta _{1}$                           & $0.805\pm 0.009$   
  & $\eta _{G}$                           & $4.65\pm 0.68$     \\ 
$\eta _{2}$                           & $3.94\pm 0.03$     
  & $\delta \left( C_{HAD}^{744}\right) $ & $0.95\pm 0.42$     \\ 
$A_{NS}$                              & $8.60\pm 0.18$     
  & $\delta \left( C_{HAD}^{770}\right) $ & $0.28\pm 0.27$     \\ 
$A_{S}$                               & $1.47\pm 0.04$     
  & $\delta \left( C_{\mu }\right) $      & $0.21\pm 0.18$     \\ 
$\eta _{S}$                           & $7.67\pm 0.13$     
  & $\delta \left( \sigma ^{\overline{\nu }}/\sigma ^{\nu }\right) $ 
                                      & $0.04\pm 0.50$ 
\end{tabular}
\end{center}
\label{tab:fitresults}
\end{table}

\end{document}